# Simultaneous observation of high order multiple quantum coherences at ultralow magnetic fields

K. Buckenmaier[a], K. Scheffler[a,b], M. Plaumann[c], P. Fehling[a], J. Bernarding[c], M. Rudolph[a,d], C. Back[d], D. Koelle[d], R. Kleiner[d], J.-B. Hövener[e] and A. N. Pravdivtsev[e]

[a] High-Field Magnetic Resonance Center, Max Planck Institute for Biological Cybernetics, Tübingen, Germany

[b] Department for Biomedical Magnetic Resonance, University of Tübingen, Germany

[c] Institute for Biometrics and Medical Informatics, Otto-von-Guericke University, Magdeburg, Germany

[d] Physikalisches Institut and Center for Quantum Science (CQ) in LISA[+], University of Tübingen, Germany

[e] Section Biomedical Imaging, Molecular Imaging North Competence Center (MOIN CC), Department of Radiology and Neuroradiology, University Medical Center Kiel, Kiel University, Kiel, Germany

## Abstract:
We present a method for the simultaneous observation of heteronuclear multi-quantum coherences (up to the 3rd order), which give an additional degree of freedom for ultralow magnetic field (ULF) MR experiments, where the chemical shift is negligible. The nonequilibrium spin state is generated by Signal Amplification By Reversible Exchange (SABRE) and detected at ULF with SQUID-based NMR. We compare the results obtained by the heteronuclei Correlated SpectroscopY (COSY) with a Flip Angle FOurier Series (FAFOS) method. COSY allows a quantitative analysis of homo- and heteronuclei quantum coherences.

**Keywords**: SABRE, SQUID, hyperpolarization, NMR, parahydrogen, ULF, $^{19}$F, $^{1}$H, multiple quantum coherence

The hyperpolarization of nuclear spins and the associated breakthroughs in physics, chemistry, biology and medicine continue to inspire the work of many scientists around the world. During the past decades, various hyperpolarization techniques have been developed [1–14] to boost the magnetic resonance (MR) signal in order to bring new applications to industry [15–17] and life-sciences [18–25]. One of these methods is Signal Amplification By Reversible Exchange (SABRE) [26–36], where parahydrogen ($p$H$_2$) is used to polarize dissolved molecules by their mutual exchange with transient complex. SABRE is unique in providing continuous hyperpolarization in the liquid state [26,37], provides high-throughput and is relatively cost efficient. Much effort is being undertaken to bring SABRE to "life"; - still, despite considerable efforts, a clean, highly polarized, highly concentrated biologically relevant contrast agent was not produced yet [38–43].

When it comes to biomedical application, usually, it is the goal to populate one dedicated spin state that boosts the MRI signal of the given nuclei.. Here, we report on an opposite effect: we discovered that it is possible to simultaneously hyperpolarize multiple, homo- and heteronuclear multi-quantum coherences, up to the 3rd order, by SABRE. This observation illustrates that the transfer of spin order from $p$H$_2$ to a substrate is still not yet completely understood. There is evidence that this transfer results not only in the substrate's magnetization, but also in the population of multiple spin orders, including homo- and heteronuclear zz-orders or singlet spin states [27,35,36,44–46].

In the following we first describe our measurement system and then turn to sample preparation and the numerical code to analyze the data. This is followed by a description of two different pulse sequences used for the experiments - a Flip Angle FOurier Series sequence (FAFOS) and a modified COrrelation SpectroscopY (COSY). These two sections also contain the results obtained and their analysis.

The effect of the hyperpolarization of multi-quantum coherences was revealed using a Superconducting QUantum Interference Device (SQUID)-based UltraLow-Field (ULF) Magnetic Resonance (MR) system, designed as a field cycling system, which operates in the range of 10 µT to 20 mT. In this range of magnetic fields SQUID-based magnetic field detectors [47] and optical magnetometers [48] become superior to conventional Faraday coils [49]. In essence, our experimental setup consisted of a coil generating the static magnetic field ($B_0$), a Helmholtz coil to excite the spins by pulses ($B_1$), a polarizing coil to generate the (elevated) field for SABRE ($B_p$) and a SQUID-based detector, which was positioned inside a low noise fibreglass dewar (**fig. 1(a)**). The polarizing coil enabled fast switching (≈ 10 ms) between $B_p$ and $B_0$ and was placed inside a three-layered mu-metal shield. For the experiments, $B_p$ was set to the optimal strength for the SABRE hyperpolarization of $^{19}$F of the ligands, 5.2 mT [35], and reduced to $B_0$ = 91.2 µT for observation (see scheme on **fig. 1 (b,c) and S1 (a)**). The setup is described in detail in Ref. [50].

Our experiments were carried out using two different samples. **Sample 1** contained 29.64 µl 3-fluoropyridine (3FPy) and 10.5 mg of the IrIMesCODCl complex [51] self-synthesized according to Ref. [52] (IMes = 1,3-bis (2,4,6-trimethylphenyl) imidazole-2-ylidene, COD = cyclooctadiene). Both substances were dissolved in 15 ml methanol. **Sample 2** was prepared using 48.76 µl ethyl 5-fluoronicotinate (EFNA) and the same amount of the IrIMesCODCl catalyst (10.5 mg) and methanol (15 ml). Results obtained with **sample 1** are reported in the main text, while those of **sample 2** are in **SM.** All samples were filled into a 2 ml sample container that was held at room temperature at the isocenter of the SQUID-based MR system under atmospheric pressure. This container was connected to a reservoir filled with the rest of the sample (see **fig. 1 (a)**). $p$H$_2$ was bubbled continuously through the sample container at a rate of ≈ 42 Scm$^3$/min. The experiment was carried out for about 8 hours until ≈ 13 ml of the solution was evaporated as monitored by MR (see **fig. S9 in SM).** The initial concentrations of the Ir-catalyst and of the fluorinated ligand were 1 mM and 23 mM, respectively.

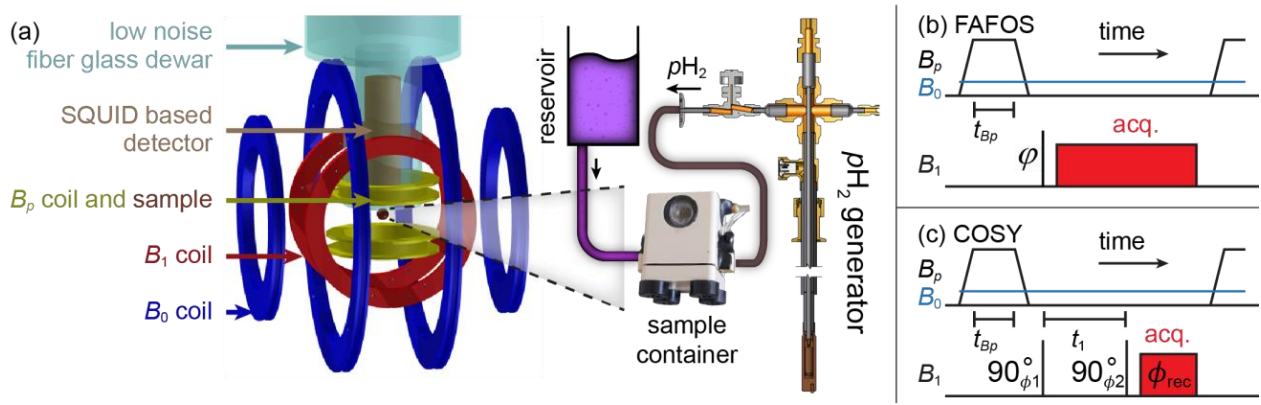

**Figure 1**. Scheme of the ULF SQUID based MR setup for the hyperpolarization of multiple quantum coherences by SABRE (a). Freshly produced $pH_2$ was supplied to the sample container inside a magnetic shield, where the field was varied between the polarizing field $B_p$ and the measurement field $B_0$ and either of two pulse sequences, FAFOS (b) or COSY (c), was used. For details, see Ref. [50].

To analyze the experimental results, density matrix simulations were performed using the code of the Magnetic resonance Open source INitiative (MOIN) [53] [27] in the following steps:

1. Setting up a the spin system of the non-polarized substrate: four $^1$H, one $^{19}$F for 3FPy or three $^1$H, one $^{19}$F for EFNA;
2. Additions of two singlet-state hydride protons ($pH_2$) to the system forming the polarized Ir – complex: IrHH-3FPy or IrHH-EFNA, where IrHH represents the hydride protons of transient Ir - complex;
3. Removal of coherences of the density matrix written in the eigenbasis of the systems' Hamiltonian at $B_p$ (polarization transfer in Ir-complex);
4. Removal of the two hydride protons from the system (dissociation of the substrate);
5. Application of the pulse sequence to the free substrate, dissolved $H_2$ and IrHH-3FPy or IrHH-EFNA complex. The Liouville von Neuman equations was used to evolve the spin system.

To investigate the effect of pulse sequences on the Ir-complex, step 4 (dissociation) was omitted. J-coupling constants were taken from literature or were estimated [35], excitation pulses were treated as ideal rotations with zero duration (hard pulses), and spin relaxation was neglected. More details and an analytical description of the sequence performance is given in supplementary materials (**SM, sections 2-4)**.

We next turn to the pulse sequences used and the results obtained with them. FAFOS (**fig. 1 (b)**) [54–56] was used to reveal hyperpolarized high-order spin states. FAFOS is a simple 1D sequence where the flip angle $\varphi$ is varied. The investigation of the components of a Fourier series along this flip angle sweep showed the presence of high-order spin states (**SM, section 4**). However, a quantitative analysis with this method proved difficult; the overall amplitudes of the first three Fourier components had similar relative intensities compared to simulations, in which higher order spin states were populated. The phase and amplitude of each peak could not be correlated and as a result, the influence of different hyperpolarized spin orders was not identified. The simulations revealed that almost perfect hard excitation pulses are necessary for such a quantitative analysis – experimentally, this was difficult to realize. Because of the limited excitation bandwidth, the excitation of the nuclei varied by 5 – 10 % (**SM section 3**).

Still, some insights from the experiment were gained. For example, we found that not only the substrate was polarized but also $H_2$ and IrHH protons [57] (**fig. 2**). Prominent features of the $^1$H-SABRE spectrum of **sample 1** were reproduced, when the spectra of the Ir-complex (IrHH-3FPy), $H_2$ and substrate were added up after weighting by 0.44 ± 0.05, 0.04 ± 0.03 and 0.52 ± 0.08, respectively (ad hoc weight factors to fit the data). Note that these factors were used for reproducing the spectrum, and are not related to the polarization levels. All components have different volume fractions and relaxation times, which were not taken into account. Also the amount of $H_2$ is difficult to predict, further exacerbating a quantitative analysis. In Ref. [35], the central, positive component of the sample spectrum was tentatively attributed to hyperpolarized methanol. In the light of our results, however, it appears more likely that it is a mix of $H_2$, 3FPy and IrHH-3FPy. The here demonstrated and theoretically explainable solution to this problem appears to be much more plausible and consequently results in the possibility to directly observe $H_2$ and IrHH signals at ULF. The sign of polarization can serve as a contrast technique at ULF. This is important, because due to the negligible chemical shift at ULF (at 91 µT the chemical shift difference of 1 ppm corresponds to only 4 mHz frequency variation), it is difficult to distinguish different chemical substances. Therefore, usually relaxation or diffusion properties are used to distinguish e.g., different types of tissue and substances [58].

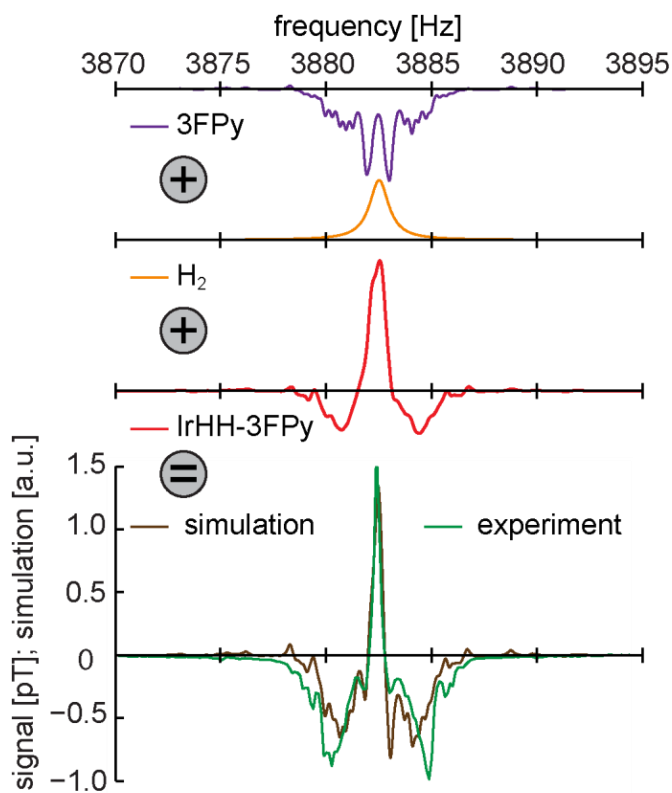

**Figure 2**. Simulated ULF SABRE $^1$H-spectra of hyperpolarized 3FPy substrate (purple), $H_2$ (orange) and IrHH-3FPy complex (red). The weighted sum of these spectra (brown) was fitted to the experimental data (green). This data indicates that hyperpolarized $H_2$ and Ir-complex were detected at ULF.

To elucidate the hyperpolarized spin states further, we modified COSY (**fig. 1 (c)**) for ULF [59,60]. After a similar preparation like for FAFOS, two $90^0$ $^1$H-$^{19}$F excitation pulses with the phases $\phi_1$ and $\phi_2$ and a variable interpulse interval, $t_1$, were applied. After the second 90º pulse, the signal was acquired with a receiver phase $\phi_{\text{rec}}$. A 2D Fourier transformation along the direct readout (→ *frequency* 1) and the indirect "$t_1$ direction" (→ *frequency* 2) resulted in a 2D spectrum. In contrast to FAFOS, many hyperpolarized QCs were identified as peaks in the *frequency 2* domain. All of these were reproduced by simulations (**fig. 3 (a,b)**). A relatively narrow bandwidth in *frequency 2* domain was chosen to obtain highly resolved 2D spectra. All essential features were reproduced by simulations (**fig. 3 (b)**).

Due to the limited spectral width (SW) of 4 kHz in the indirect dimension, the signals of the quantum coherences fold into the bandwidth between –2 kHz and 2 kHz (see **fig. 3** and **tab. S5**). Such frequencies can be easily obtained using the product operator formalism [61] (see **section 7**, **SM**). A broader SW of 40 kHz at the given field would be necessary to prevent folding of the QCs to lower frequencies. Such a high SW would expand the measurement time tremendously and would not be practical. Altogether, theoretically a maximum of 21 hyperpolarized signals at different resonance frequencies can be obtained for 3FPy with COSY without phase cycling (see **tab. S5**). By chance, at the given magnetic field, $B_0$ = 91.18 µT, and SW = 4 kHz, six pairs of peaks have exhibited the same (aliased) frequencies. Therefore, only 15 peaks were clearly separable in the simulations. Experimentally only 11 were obtained (**fig. 3 (a)** and **S5, S6**). These peaks belonged to QCs up to the 3rd order. The signal intensity of higher order QCs was below the noise level of the setup. Only two peaks were "diagonal" COSY peaks, which are the result of the -1 quantum coherence evolution during the $t_1$ time interval and acquisition block (indicated by red dashed lines on **fig. 3**). All other peaks are "cross-peaks".

The COSY experiment with phase cycling solves the problem of overlying aliased signals. For the selective observation of multi-quantum coherences the experiment was repeated four times with the following phases of two excitation pulses (**fig. 1 (c)**): $\phi_1$ = x, y, -x, -y, $\phi_2$ = 4(x). Then, four different $\phi_{rec}$ cycles were used to select different orders, $p_1$, of quantum coherences during the $t_1$ interval: (A) $\phi_{rec}$ = x ,-y, -x, y selects $p_1$ = 1+4n; (B) $\phi_{rec}$ = x, y, -x, -y selects $p_1$ = -1+4n; (C) $\phi_{rec}$ = x, -x ,x, -x selects $p_1$ = 2+4n; (D) $\phi_{rec}$ = 4(x) selects $p_1$ = 4n with $n$ being an integer number (see **tab. S4**). There are 11 multiple quantum coherences for a 5 spin-1/2 system: 0, $\pm 1$, $\pm 2$, $\pm 3$, $\pm 4$ and $\pm 5$. And all of them can be excited with the first $90^0$ pulse (**eq S3**, **SM**) if appropriate spin orders are populated. In **fig. 3 (c)** two out of four coherence selective COSY spectra are shown (additional experimental COSY results are given in **SM** (**fig. S5-S8**).

Therefore, with COSY at ULF various coherences can be measured directly at different resonance frequencies (i), there is no need for a very precise extraction of small signals from the one big signal (like in FAFOS or any 1D experiments) that comprises different spin orders (ii) and different phase cycling schemes select different coherences (iii). Although QCs up to the fifth order could theoretically be measured (up to the fourth order for EFNA), we could detect and assign only QCs up to the third order. As the simulations showed, the signal intensities of higher order QCs is dropping rapidly and are below the noise level of the setup.

To demonstrate that higher order quantum coherences in COSY spectra are the result of the population of higher order spin states and not a coherent evolution of magnetization during the COSY sequence we performed additional simulations (see **fig. 4**). A COSY spectrum of a system with only longitudinal initial polarization showed quantum coherences up to the first order.

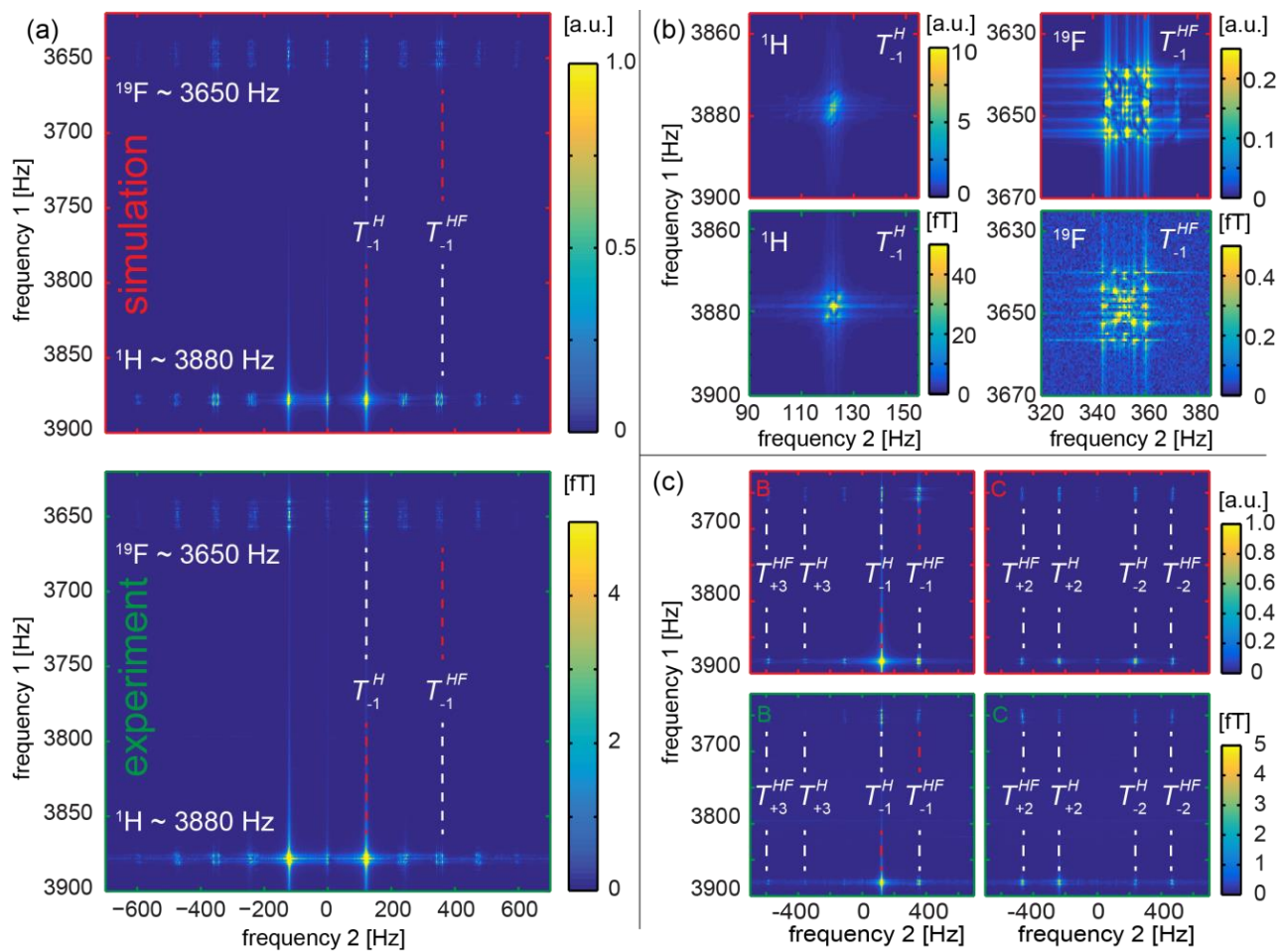

**Figure 3.** Simultaneous hyperpolarization and separation of multiple QCs using COSY: Simulated (green frames) and experimental (red frames) $^1$H-spectra of 3FPy (a), magnified views (b), and different phase-cycling variants of the sequence (c), where B and C corresponds to phase cycling schemes given in **tab. S4**. In $T_y^x$, $x$ indicates the nuclei involved, and $y$ the order of the QC (**tab. S5**). Red and white dashed lines mark the diagonal and off-diagonal peaks, respectively. Up to third order coherences were observed, and any experimental details were reproduced by simulations (b), despite the fact that some J-couplings had to be guessed (**tab. S1**).

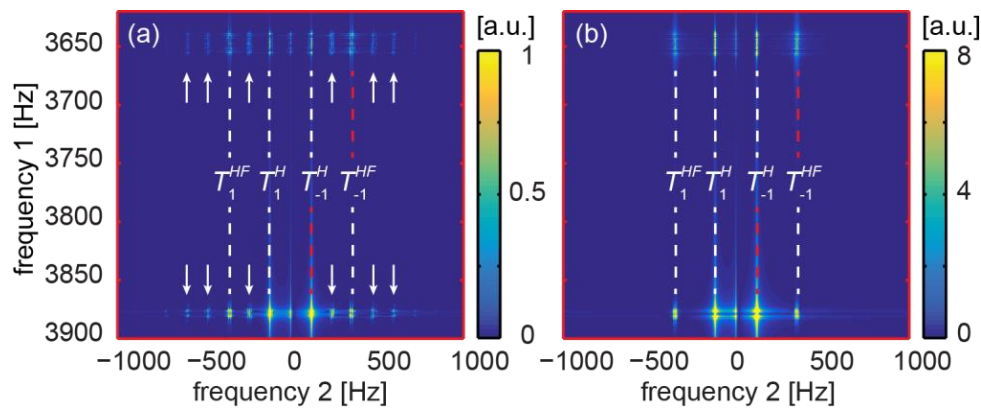

**Figure 4.** Simulated ULF COSY amplitude spectra of 3FPy after SABRE hyperpolarization at $B_p$ = 5.2 mT (a) or after artificial hyperpolarization of longitudinal states alone (b). Note that in case of SABRE higher order quantum coherences appear (up to the 5th order for 3FPy marked by white arrows). The COSY spectrum with initial longitudinal magnetization shows only first order (and zero H-H order) quantum coherences. Also note that the shapes of the peaks are different. $B_0$ was 91.1 µT and J-coupling constants for 3FPy are listed in **tab. S1**.

To summarize: in this contribution, we demonstrated the hyperpolarization of heteronuclear, multiple QCs using a SQUID based ULF MR and SABRE. We developed two different approaches to analyze hyperpoloriyed multi-spin system: FAFOS and COSY. As a result, we detected QCs up to the third order that demonstrates that multi-spin orders are populated in SABRE too. Even so there were already quantitative effects of QCs measured in high field NMR [46], to our knowledge the experiments presented here showed for the first time that multiple quantum coherences between protons and heteronuclei can be hyperpolarized by SABRE at low magnetic fields: one of the reasons for a relatively low efficiency of SABRE on substrates with many coupled spins. The observation of the polarization by COSY helped to improve the understanding of the SABRE polarization transfer on a physical level and showed, that the spin alignment distribution process can be described via the full density matrix approach used here. Moreover, COSY at ULFs is also one of the few experiments, where indirectly multiple quantum coherences of liquid samples in the range from -3 to +3 are observed within a single experiment.

We believe that this contribution is an important experimental confirmation of the commonly known and underestimated evidence that at low magnetic field polarization distributes among all strongly-coupled spins even if the direct source-target nuclear spin coupling is small. Moreover, the demonstration that multiple quantum coherences are hyperpolarized with SABRE is an important information for continuous improvement of SABRE performance but it may be even more important for some exotic methods such as ultra low field magnetometers [62] and the Radiowave Amplification by Stimulated Emission of Radiation RASER pioneered by Appelt et al. [13].

## Additional materials

### Supporting Information

NMR parameters of 3FPy and EFNA, additional materials to ULF SABRE FAFOS and COSY, measurement of signal stability during the long lasting experiemnts, evaluation of QCs frequencies and used phase cycling schemes (.PDF), Matlab simulation source code (.m) are available as supplementary.

**Author Contributions**

All authors have given approval to the final version of the manuscript.

# Acknowledgement

A.P. and J.-B.H. acknowledge support by the Emmy Noether Program of the DFG (HO 4604/2-1), the Cluster of Excellence "Inflammation at Interfaces" (EXC306). Kiel University and the Medical Faculty are acknowledged for supporting the Molecular Imaging North Competence Center (MOIN CC) as a core facility for imaging in vivo. MOIN CC was founded by a grant of the European Regional Development Fund (ERDF) and the Zukunftsprogramm Wirtschaft of Schleswig-Holstein (Project no. 122-09-053). This work was supported by the Deutsche Forschungsgemeinschaft (BE 1824/12-1). K.B. thanks Jonas Bause for fruitful discussions.

# SUPPORTING INFORMATION

## Contents

# 1. NMR parameters of 3FPy and EFNA

**Table S1**. J-coupling constants in Hz of 3FPy and estimated coupling of 3FPy with IrHH protons (IrH[a] and IrH[b]). $T_1$-relaxation times are estimated or measured at high magnetic field and used for simulations. For simplicity, we assume that the chemical shift of 3FPy in the Ir-complex and in the solution does not change. Gyromagnetic ratios are taken from Bruker Almanac 2011 [1].

| 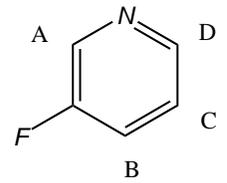 | IrH[a] | IrH[b] | A | B | C | D | F |
|---|---|---|---|---|---|---|---|
| IrH[b] | -7 | | | | | | |
| A | 1.2 | 0 | | | | | |
| B | 0 | 0 | 2.8 | | | | |
| C | 0 | 0 | 0.7 | 8.6 | | | |
| D | 0 | 0.5 | 0 | 1.2 | 4.6 | | |
| F | 1 | 0 | 1 | 8.75 | 4.9 | 1.7 | |
| δ, ppm | -22 | -22 | 8.47 | 7.65 | 7.49 | 8.40 | -127.72 |

**Table S2**. J-coupling constants in Hz of EFNA and estimated coupling of FNA with IrHH protons (IrH[a] and IrH[b]). $T_1$-relaxation times are estimated or measured at high magnetic field and used for simulations. For simplicity, we assume that the chemical shift of EFNA in complex and in bulk does not change.

| 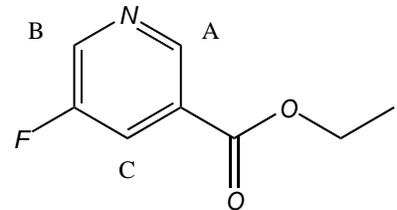 | IrH[a] | IrH[b] | A | B | C | F |
|---|---|---|---|---|---|---|
| IrH[b] | -7 | | | | | |
| A | 1.2 | 0 | | | | |
| B | 0 | 0.5 | - | | | |
| C | 0 | 0 | 1.62 | 2.93 | | |
| F | 1 | 0 | 1.60 | 0.75 | 8.76 | |
| δ, ppm | -22 | -22 | 8.99 | 8.70 | 8.14 | -127.72 |

## 2. Pulse Sequence parameters

The sequence parameters are listed in **tab. S3**. The meaning of the parameters is illustrated in **fig. S1**. During the whole sequence, $pH_2$ was bubbled through the sample with a rate of ≈ 42 scm$^3$/min and $B_0$ was kept at 91.18 µT. Bp was switched between 0 and 5.2 mT. In **tab. S3** also the frequencies of the $B_1$ excitation pulses are listed. For determining the $B_0$ magnetic field strength and corresponding $^1$H NMR frequency, the resonance frequency of the central (methanol) peak of the $^1$H signal was used.

**FAFOS.** For the FAFOS sequence, the flip angle was varied from 0° to 360° with a step size of approx. 3.6°. Altogether $N$ = 101 1D spectra with variable $\varphi$ were acquired. After that, a Fourier series along the $\varphi$-dimension was calculated (eq 4). It has to be mentioned that the flip angle was determined from a calibration measurement for a $B_1$ pulse exciting only $^1$H or $^{19}$F. The excitation pulse for both nuclei was difficult to calibrate for two reasons. The first is the influence of the multiple quantum coherences. And the second is a crosstalk due to a not small enough excitation bandwidth of the two frequency components of the pulse. The total time for the experiment was ~ 1.5 h.

**COSY.** For the COSY sequences, the hyperpolarization time was kept as short as possible in order to reduce TR, by the cost of a lower signal intensity. A short TR enables more measurement steps, as a result, higher spectral resolution and broader spectral width in the indirect $t_1$-encoding dimension is achieved.

**COSY 3FPy**. The measurement of the COSY spectrum for 3FPy without phase cycling (PC) was separated into 18 blocks. Each block contains 400 of constantly incremented $t_1$ steps with $\Delta t_1$ = 0.25 ms, which corresponds to a spectral width (SW) of 4 kHz in the indirect $t_1$-dimension of the COSY spectrum. An overall record of 18 blocks allowed us to vary $t_1$ from 20 ms up to 1820 ms, which corresponds to a spectral resolution of 0.6 Hz. After each block, a control spectrum was acquired.

**COSY EFNA**. The measurement of the COSY spectrum for EFNA without PC was separated into 20 blocks. Each block contains 200 of constantly incremented $t_1$ steps with $\Delta t_1$ = 0.5 ms, which corresponds to SW = 2 kHz in the indirect $t_1$-dimension of the COSY spectrum. An overall record of 20 blocks allowed us to vary $t_1$ from 20 ms up to 2020 ms, which corresponds to a spectral resolution of 0.5 Hz. After each block, a control spectrum was acquired.

**COSY PC 3FPy**. The measurement of the COSY spectrum with PC for 3FPy was separated into 19 blocks. Each block contains 80 of constantly incremented $t_1$ time steps. As for the previous measurement, after each block, a control measurement was performed. Since each $t_1$ time step needs 4 phase cycling steps the SW and resolution were lower as for the measurement without phase cycling. $\Delta t_1$ was set to 0.5 ms resulting in SW = 2 kHz and a spectral resolution of 1.3 Hz.

**COSY PC EFNA**. The measurement of the COSY spectrum with PC for EFNA was separated into 14 blocks. Each block contains 80 of constantly incremented $t_1$ time steps. After each block, a control measurement was performed. $\Delta t_1$ was set to 0.5 ms resulting in SW = 2 kHz and a spectral resolution of 1.8 Hz.

**Table S3.** Sequence parameters for the FAFOS and the COSY sequences for the measurements shown. The parameters are illustrated in fig. SISEQ.

|  | 3FPy | | | EFNA | | |
|---|---|---|---|---|---|---|
|  | FAFOS | COSY | COSY PC | FAFOS | COSY | COSY PC |
| $t_{ramp}$ [ms] | 30 | 30 | 30 | 30 | 30 | 30 |
| $t_{Bp}$ [s] | 4 | 2 | 2 | 4 | 2.5 | 2.5 |
| $t_w$ [ms] | 11 | 11 | 11 | 11 | 11 | 11 |
| $t_{B1}$ [ms] | 15 | 15 | 15 | 20 | 15 | 15 |
| $t_{B1,w}$ [ms] | 7 | 7 | 4 | 7 | 7 | 7 |
| $t_{acq}$ [s] | 8 | 2 | 2 | 8 | 2 | 2 |
| TR [ms] | 12150 | 4300 – 5900 | 4169 – 4829 | 12160 | 4572 – 6572 | 4572 – 5132 |
| $t_1$ [ms] | — | 20 – 1720 | 20 – 780 | — | 20 – 2020 | 20 – 580 |
| steps | 141 | 6801 | 1521 | 241 | 4001 | 1121 |
| avg. | 2 | 1 | 4 | 2 | 1 | 4 |
| $B_1$ freq. $^1$H [Hz] | 3880 | 3875 | 3875 | 3880 | 3880 | 3880 |
| $B_1$ freq. $^{19}$F [Hz] | 3650 | 3650 | 3650 | 3650 | 3650 | 3650 |
| $^1$H NMR freq. [Hz] | 3882.5 | 3877.9 | 3881.1 | 3882.0 | 3881.9 | 3882.0 |

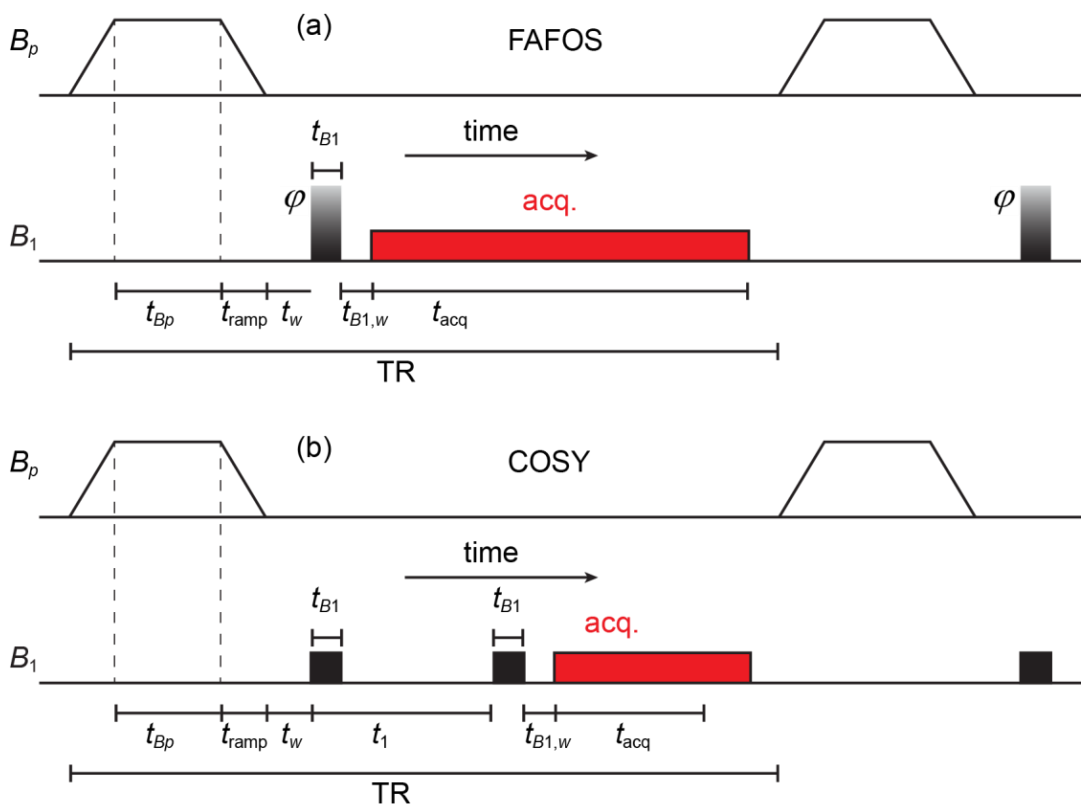

**Figure S1.** Illustration of sequence parameters for (a) the FAFOS and (b) the COSY sequences.

## 3. $^1$H-$^{19}$F-Double frequency excitation pulse

The $B_1$ excitation pulse can be described by the expression:

$$f(t) = (\sin(\omega_{1H}t) + \sin(\omega_{19F}t)) \cdot \text{sinc}(4 \cdot (t - t_{end}/2)/t_{end}),$$

Here $t_{end}$ is the length of the total pulse and $\omega_{1H}$ and $\omega_{19F}$ are the excitation frequencies. An excitation pulse used for the experiments is shown in **fig. S2 (a)**. **Figure S2 (b)** shows the single sided Fourier spectrum $F(\omega)$ of the pulse.

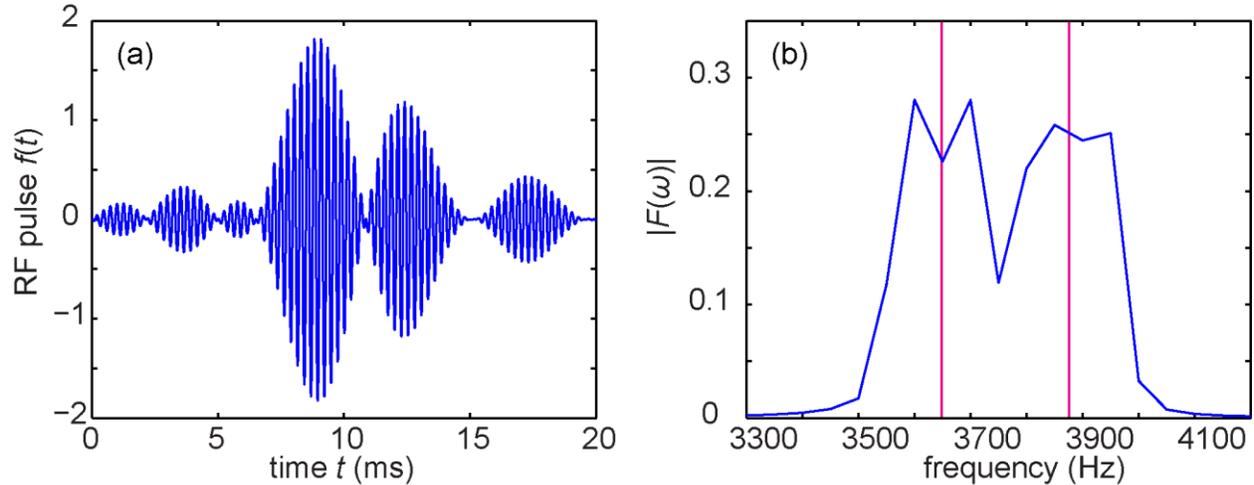

**Figure S2**. Sinc shaped modulated excitation pulse (a) and the corresponding amplitude spectrum (b). The excitation profile allows simultaneous excitation at two frequencies. At $B_0$ = 91.18 µT resonance frequencies of $^1$H and $^{19}$F are ~ 3880 Hz and ~ 3650 Hz respectively (marked on (b)). Similar excitation profiles were used in the experiments shown in the manuscript. Note that because the two excitation regions are overlapping and have different intensities the experimental excitation flip angle could deviate from the nominal value by 5 – 10%.

## 4. FAFOS: Additional results

If the full density matrix approach is used to calculate the spin state of a SABRE system at ULF [2,3], it is easy to show that not only longitudinal magnetization but also collective spin orders with total zero-quantum coherence are populated.

For an *N* spin-½ system, the longitudinal magnetization is represented by *N* spin operators. Since we are interested e.g., in 3FPy, hereafter a five spin system with four longitudinal $^1$H operators $\hat{T}_Z = \hat{I}_Z, \hat{S}_Z, \hat{R}_Z, \hat{K}_Z$ and one $^{19}$F operator $\hat{T}_Z = \hat{F}_Z$ is discussed. These operators represent the spins of the nuclei A, B, C, D and F of 3FPy (**tab. S1**, **SI**).

The collective spin orders of a total zero-quantum coherence are represented by ten $\hat{T}_{ZZ}$ terms (e.g., $\hat{I}_Z\hat{R}_Z$), ten $\hat{T}_{3Z}$, terms (e.g., $\hat{I}_Z\hat{S}_Z\hat{R}_Z$), five $\hat{T}_{4Z}$ terms (e.g., $\hat{I}_Z\hat{S}_Z\hat{R}_Z\hat{T}_Z$), one $\hat{T}_{5Z} = \hat{I}_Z\hat{S}_Z\hat{R}_Z\hat{T}_Z\hat{F}_Z$ term and more other spin orders that are combinations of zero-quantum coherences with various $\hat{T}_Z$ terms (e.g., $\hat{I}_X\hat{S}_X\hat{F}_Z + \hat{I}_Y\hat{S}_Y\hat{F}_Z$). It is very difficult to distinguish between different spin orders by 1D high resolution NMR spectroscopy. At ULF the conditions are even more complicated, because of a negligible chemical shift all nuclei of one type have the same resonance frequency.

The first approach we discussed in the main text is FAFOS. When a $B_1$ pulse is applied to a system that is described by one of the $\hat{T}_Z$ terms the observed signal is proportional to $\sin(\varphi)$ and the "−1" quantum coherence ($\hat{T}_{-1}$) is observable. When a system is described by other spin orders, the observed signal amplitudes are proportional to the functions:

$$\hat{T}_Z \xrightarrow{\varphi} \sin(\varphi)\,\hat{T}_{-1}$$
$$\hat{T}_{ZZ} \xrightarrow{\varphi} \sin(\varphi)\cos(\varphi)\,\hat{T}_{-1} = \tfrac{1}{2}\sin(2\varphi)\,\hat{T}_{-1}$$
$$\hat{T}_{3Z} \xrightarrow{\varphi} \sin(\varphi)\cos^2(\varphi)\,\hat{T}_{-1} =$$
$$= \tfrac{1}{4}(\sin(\varphi) + \sin(3\varphi))\hat{T}_{-1}$$
$$\hat{T}_{4Z} \xrightarrow{\varphi} \sin(\varphi)\cos^3(\varphi)\,\hat{T}_{-1} =$$
$$= \tfrac{1}{8}(2\sin(2\varphi) + \sin(4\varphi))\hat{T}_{-1}$$
$$\hat{T}_{5Z} \xrightarrow{\varphi} \sin(\varphi)\cos^4(\varphi)\,\hat{T}_{-1} =$$
$$= \tfrac{1}{16}(2\sin(\varphi) + 3\sin(3\varphi) + \sin(5\varphi))\hat{T}_{-1}$$

(**eq S1**)

The amplitude of the different $\sin(n\varphi)$ terms can be obtained by evaluating the Fourier coefficient series (FCs), $c_k(\omega)$, of the spectra $S(\omega, \varphi)$ in the direction of the flip angle sweep with

$$c_k(\omega) = \sum_{j=1}^{L} S(\omega, \varphi_j) \sin(k\varphi_j) \quad (\textbf{eq S2})$$

where $k$ is a positive integer numbers, $\varphi_j = \frac{2\pi}{L}j$ is the flip angle and *L* is given by $\varphi_L = 2\pi$.

Hence, this method perfectly separates $\hat{T}_Z$ from $\hat{T}_{ZZ}$, if no other terms are present. However, all odd and even $\hat{T}_{nZ}$ terms have some common $\sin(n\varphi)$ functions, e.g., for $\hat{T}_Z$, $\hat{T}_{3Z}$ and $\hat{T}_{5Z}$ it is $\sin(\varphi)$. Therefore, a simple Fourier series along the flip angle sweep used before will not separate different spin orders and additional post processing is necessary. Although this method can be used for a demonstration of the presence of higher spin orders, a quantitative analysis is difficult. The presence of zero-quantum coherences complicates the situation and is not described by **eq. 1**.

Calculations predict that at least the first three FCs should be observable. However, it is challenging experimentally to obtain higher harmonics, because a higher order series has a lower intensity, all FCs contribute to the one signal at a given frequency and polarization level deviate a little with time. Especially the contribution of non-perfect excitation pulses makes a quantitative analysis nearly impossible (**fig. S2**).

The comparison of experimental and simulated FCs for $^1$H and $^{19}$F is shown in **fig. S3** and **S4**. The experiments showed, that a quantitative analysis is difficult with this method; the overall amplitudes of the first three FCs in experiment and in simulation for free substrate have similar relative intensities. The phase and amplitude of each peak could not be correlated and as a result, the influence of different hyperpolarized spin orders was not identified. This was the case for both substrates, 3FPy and EFNA (**fig. S3,S4**).

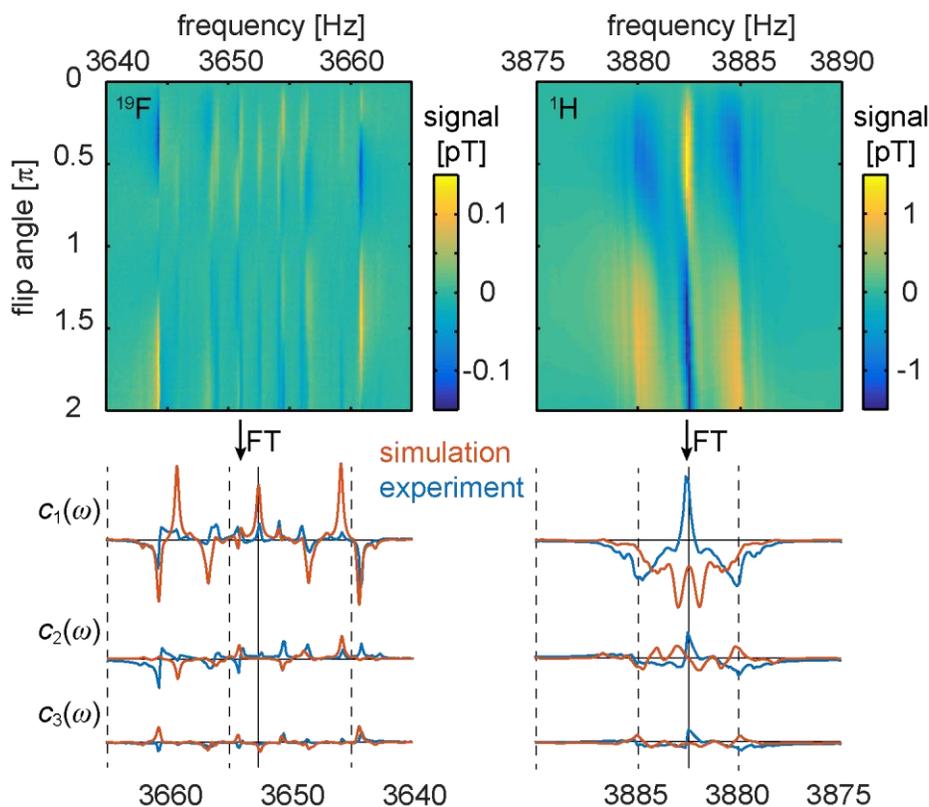

**Figure S3**. ULF SABRE FAFOS spectra $S(\omega, \varphi)$ (upper part) and comparison of experimental (blue lines) and simulated (red lines) FCs, $c_{1,2,3}(\omega)$, (lower part) for $^{19}$F (left side) and $^1$H of 3FPy. The black symmetry line indicates the center frequency of the $^{19}$F and $^1$H signal. J-coupling constants for 3FPy are listed in **tab. S1**.

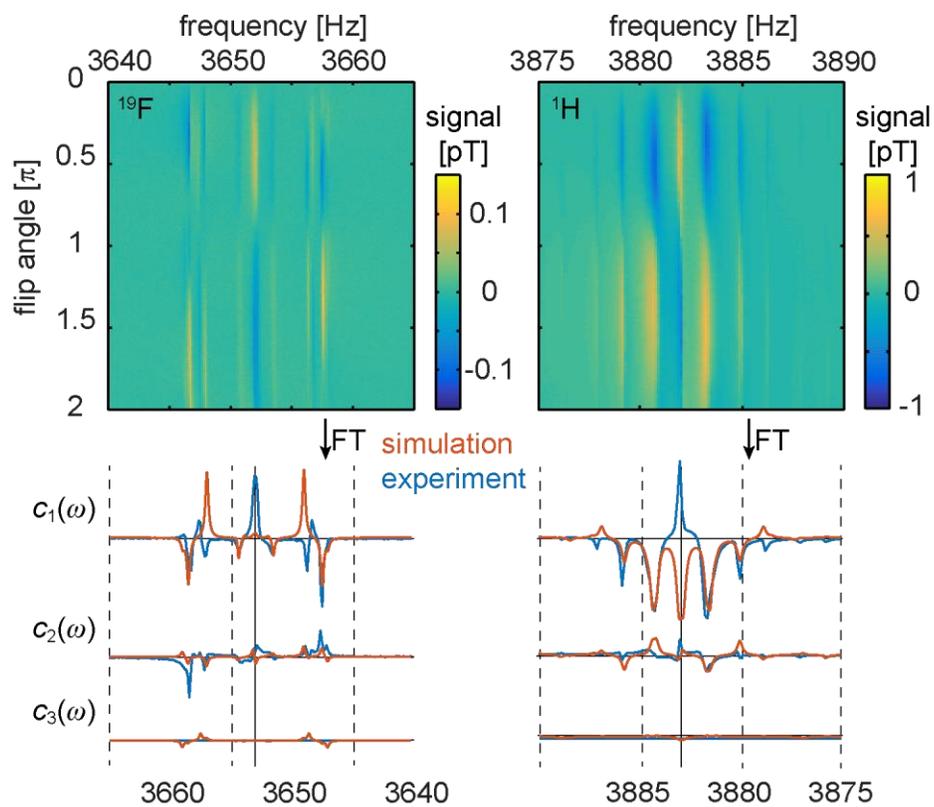

**Figure S4**. ULF SABRE FAFOS spectra (upper part) and comparison of experimental (blue lines) and simulated (red lines) FCs, $c_{1,2,3}(\omega)$, (lower part) for $^{19}F$ (left side) and $^{1}H$ of EFNA. The black symmetry line indicates the center frequency of the $^{19}F$ and $^{1}H$ signal. The simulation parameters are $B_0$ = 91.1756 µT, $B_p$ = 5.2 mT, J-coupling constants for 3FPy are listed in **tab. S2**.

## 5. COSY: Additional results

An alternative method to FAFOS discussed in the main text is based on a conventional COSY experiment. After the first $90^0_{\phi 1}$ pulse the $\hat{T}_Z$ terms transfer to the combination of single quantum coherences $\hat{T}_{\pm 1}$, i.e., $\hat{I}_Z \xrightarrow{90^o_Y} \hat{I}_X = (\hat{I}_+ + \hat{I}_-)/2$, or more generally, $\hat{T}_Z \xrightarrow{90^o} \hat{T}_{\pm 1}$. The transformation of other collective spin orders discussed can be shown in the same fashion and results in:

$$\hat{T}_Z \xrightarrow{90^o} \hat{T}_{\pm 1}$$
$$\hat{T}_{ZZ} \xrightarrow{90^o} \hat{T}_{\pm 2}, \hat{T}_0$$
$$\hat{T}_{3Z} \xrightarrow{90^o} \hat{T}_{\pm 3}, \hat{T}_{\pm 1} \quad \textbf{(eq S3)}$$
$$\hat{T}_{4Z} \xrightarrow{90^o} \hat{T}_{\pm 4}, \hat{T}_{\pm 2}, \hat{T}_0$$
$$\hat{T}_{5Z} \xrightarrow{90^o} \hat{T}_{\pm 5}, \hat{T}_{\pm 3}, \hat{T}_{\pm 1}$$

As for the FAFOS method, the odd and even terms here have common resulting coherences. Hence, although COSY spectroscopy at ULF cannot directly separate different $\hat{T}_{nZ}$ terms, it allows a selective observation of high order quantum coherences (see below).

A possible phase cycle of the COSY experiment (**fig. S2(b)**) is as follows: $\phi_1$ = x, y, -x, -y, $\phi_2$ = 4(x), $\phi_{rec}$ = x, -y, -x, y. The variation of $\phi_{rec}$ – cycle results in 4 different COSY spectra (**tab. S4, A-D**). At high magnetic field, usually one out of the four phase cycling schemes is used to select the single quantum coherence, $p_1 = -1$, during the interpulse delay, $t_1$. However, other coherences, namely $p_1 = -1 + 4n = -5, -1, 3, \ldots$ with $n$ being an integer number, are also selected but usually are neglected.

All four variants of the recording phase, $\phi_{\text{rec}}$, cycling together with coherence selection pathways are listed in **tab. S4**. All together there are 11 multiple quantum coherences for a 5 spin-1/2 system: 0, $\pm 1$, $\pm 2$, $\pm 3$, $\pm 4$ and $\pm 5$. And all of them can be excited with the first $90^0_{\phi 1}$ pulse (**eq S3**) if appropriate spin orders are populated.

**Table S4**. Four COSY multiquantum coherence selection phase cycling schemes: (A) $p_1 = -3, 1, 5$, (B) $p_1 = -5, -1, 3$, (C) $p_1 = \pm 2$, (D) $p_1 = 0, \pm 4$. $\phi_{\text{rec}}$ coincides with the total phase of the pulse sequences and provides a selection of the $p_1$ quantum coherence.

| Case / step | $\phi_1$ | Phase for $p_1 - p_0$ | $\phi_2$ | Phase for $p_2 - p_1$ | Total phase $\phi_{rec}$ |
|---|---|---|---|---|---|
| **A** | $p_1 = -3, 1, 5;\ 1 + 4n$ | | | | |
| 1 | 0° | 0° | 0° | 0° | 0° |
| 2 | 90° | 270° | 0° | 0° | 270° |
| 3 | 180° | 180° | 0° | 0° | 180° |
| 4 | 270° | 90° | 0° | 0° | 90° |
| **B** | $p_1 = -5, -1, 3;\ -1 + 4n$ | | | | |
| 1 | 0° | 0° | 0° | 0° | 0° |
| 2 | 90° | 90° | 0° | 0° | 90° |
| 3 | 180° | 180° | 0° | 0° | 180° |
| 4 | 270° | 270° | 0° | 0° | 270° |
| **C** | $p_1 = \pm 2;\ 2 + 4n$ | | | | |
| 1 | 0° | 0° | 0° | 0° | 0° |
| 2 | 90° | 180° | 0° | 0° | 180° |
| 3 | 180° | 0° | 0° | 0° | 0° |
| 4 | 270° | 180° | 0° | 0° | 180° |
| **D** | $p_1 = 0, \pm 4;\ 4n$ | | | | |
| 1 | 0° | 0° | 0° | 0° | 0° |
| 2 | 90° | 0° | 0° | 0° | 0° |
| 3 | 180° | 0° | 0° | 0° | 0° |
| 4 | 270° | 0° | 0° | 0° | 0° |

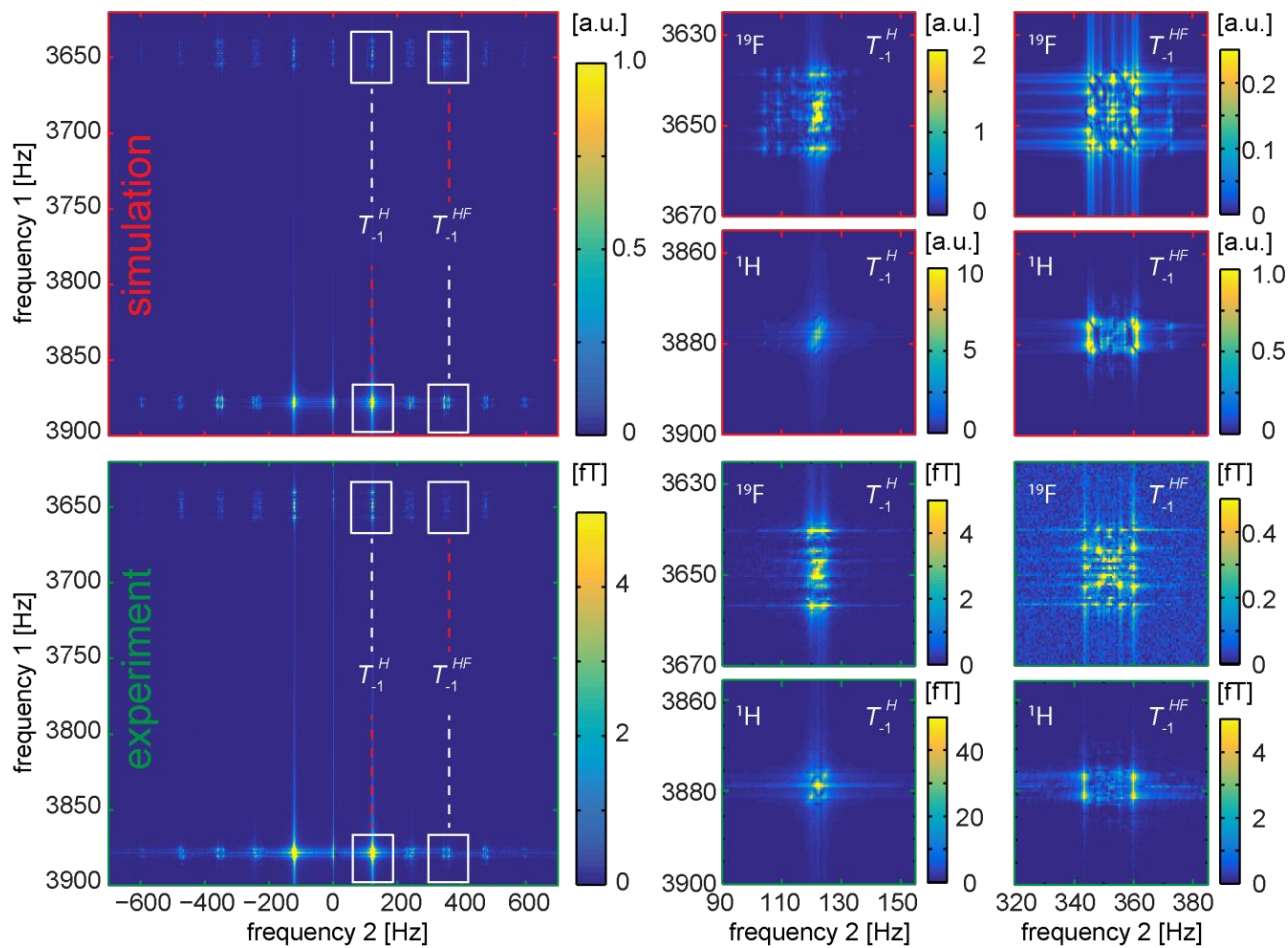

**Figure S5.** Hyperpolarization of quantum coherences up to third order: Simulated (top) and experimental (bottom) ULF SABRE COSY amplitude spectra of 3FPy. On the right, the zoomed out $T^{H}_{-1}$ and $T^{HF}_{-1}$ QCs measured by $^1$H and $^{19}$F are shown (indicated by the white rectangles). Quantum coherences up to the 3rd level are visible. J-coupling constants for 3FPy are listed in **tab. S1**.

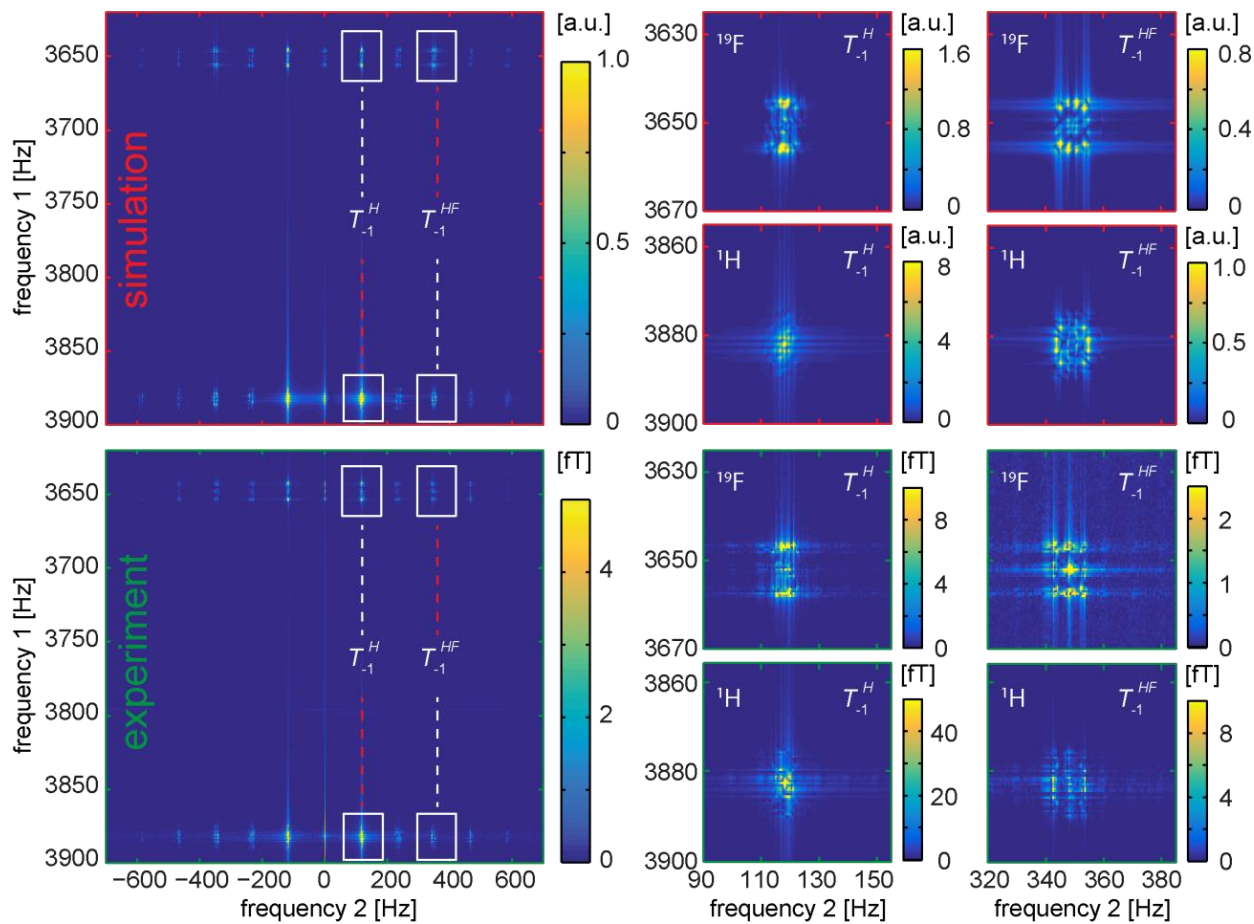

**Figure S6.** Experimental (lower half) and simulated (upper half) ULF SABRE COSY amplitude spectra of EFNA. On the right, the zoomed out $T_{-1}^{H}$ and $T_{-1}^{HF}$ QCs measured by $^1$H and $^{19}$F are shown (indicated by the white rectangles). The red and white dashed lines mark the diagonal and off-diagonal peaks respectively. The simulation parameters are $B_0$ = 91.175 µT, $B_p$ = 5.2 mT, J-coupling constants for EFNA are listed in **tab. S2**.

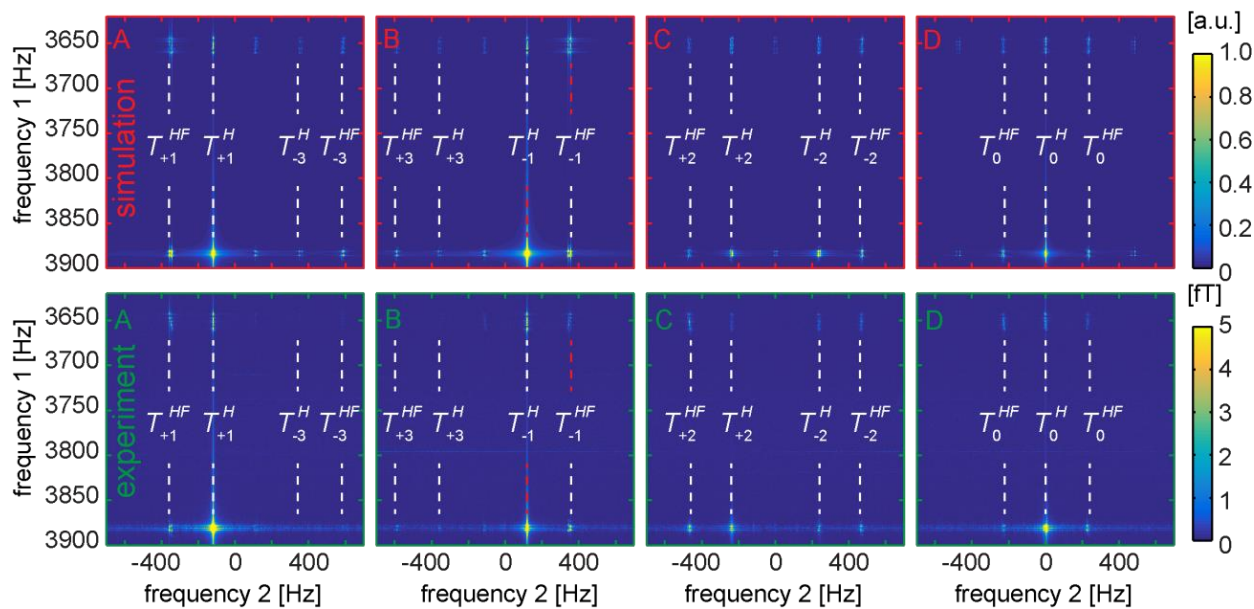

**Figure S7.** Simulated (top) and experimental (bottom) 3FPy COSY amplitude spectra obtained with four different phase alternating methods (here A-D corresponds to phase cycling schemes given in **tab. S4**). The QCs, $T$, are assigned using **tab. S5**. The red and white dashed lines mark the diagonal and off-diagonal peaks, respectively. The simulation parameters are $B_0$ = 91.175 µT, $B_p$ = 5.2 mT, J-coupling constants for 3FPy are listed in **tab. S1**.

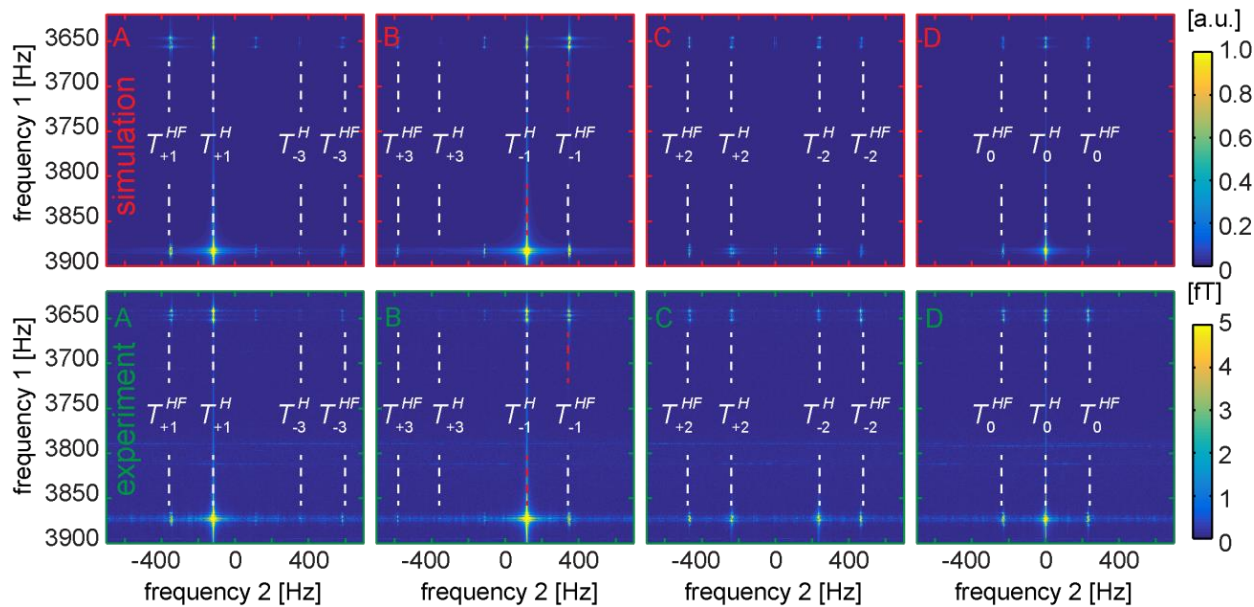

**Figure S8.** Simulated (top) and experimental (bottom) EFNA COSY amplitude spectra obtained with four different phase alternating methods (here A-D corresponds to phase cycling schemes given in **tab. S4**). The QCs, T, are assigned using **tab. S5**. The red and white dashed lines mark the diagonal and off-diagonal peaks, respectively. The simulation parameters are $B_0$ = 91.175 µT, $B_p$ = 5.2 mT, J-coupling constants for EFNA are listed in **tab. S2**.

## 6. Stability of the SABRE reaction

During all measurements control spectra were acquired. The area under the absolute value of the $^1$H peaks was integrated as a reference value for the SABRE enhancement factor. The time course is shown on **fig. S9**. The drop off at the end of each curve indicates that the level of the sample container was lowered and that the reservoir was empty.

The sequence parameters for all spectra were the same. The hyperpolarization time $t_{Bp}$ was set to 8 s and $t_{acq}$ to 8 s. The other parameters were similar to the parameters listed in **tab. S3**.

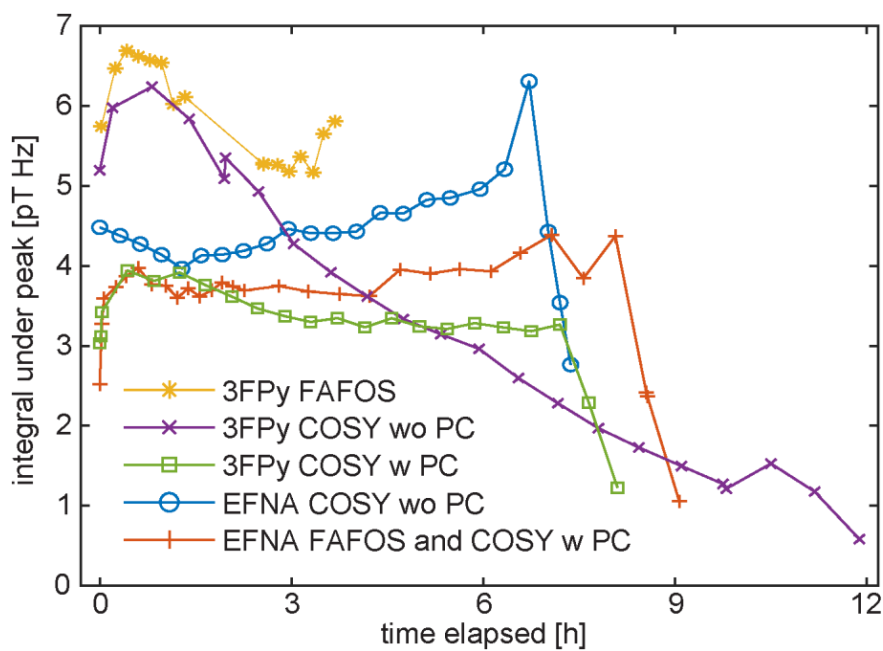

**Figure S9.** Integral under the absolute value of the $^1$H peaks as a function of time during the measurements.

## 7. COSY resonance frequencies

Spin rotation under the action of an external static magnetic field is given by the following two equations [4]:

$$\hat{I}_X \xrightarrow{-\tau\omega\hat{I}_Z} \hat{I}_X \cos(\omega\tau) - \hat{I}_Y \sin(\omega\tau)$$

$$\hat{I}_Y \xrightarrow{-\tau\omega\hat{I}_Z} \hat{I}_Y \cos(\omega\tau) + \hat{I}_X \sin(\omega\tau)$$

Since $\pm 1$ QCs, $\hat{T}_\pm$, equal to $\hat{I}_\pm = \hat{I}_X \pm i\hat{I}_Y$. Then their evolution is described as follows:

$$\hat{I}_\pm \xrightarrow{-\tau\omega\hat{I}_Z} \hat{I}_\pm e^{\pm i\tau\omega} \text{ (eq S4)}$$

From this, it follows that $T_{\pm 1}^X$ QCs evolve with $\pm\omega_X$ frequencies respectively. Using **eq S4** one can obtain evolution frequencies of all other QCs that are listed in **tab. S5**.

**Example 1**. $T_0^{HF}$ QCs are a product of one $T_{+1}^H$ and one $T_{-1}^F$ or $T_{-1}^H$ and one $T_{+1}^F$, therefore according to eq S1, $T_0^{HF}$ evolves with the frequencies $\pm(\omega_{1H} - \omega_{19F})$ accordingly.

**Example 2**. $T_{+5}^{HF}$ QC is a product of four $T_{+1}^H$ QCs and one $T_{+1}^F$, therefore according to eq S1 $T_{+5}^{HF}$ evolves with the frequency $+(4\omega_{1H} + \omega_{19F})$.

**Example 3**. $T_{+3}^{H-F}$ QC is a product of four $T_{+1}^H$ QCs and one $T_{-1}^F$, therefore according to eq S1 $T_{+5}^{HF}$ evolves with the frequency $+(4\omega_{1H} - \omega_{19F})$.

Note that states with different sign of coherences of individual spins, $T_n^{H-F}$, are not discussed in the main text because they comprise of two more spins than corresponding $T_n^{HF}$ coherence. This results in rapid decay of signal.

**Table S5**. Frequencies of multiple quantum coherences from -5 to +5 for a five spin ½ system with four protons and a single $^{19}F$ nucleus (3FPy). Upper indexes indicate that only protons (H superscript) or also $^{19}F$ (HF superscript) comprises the coherence. Aliased resonance frequencies are given here for a SW of 2 kHz or 4 kHz. Note that coherences $T_n^{H-F}$ are not discussed in the main text because they are expected to have much lower signal intensities than corresponding $T_n^{HF}$.

| QC | Frequency ($B_0$=91.18 µT), Hz | Aliased res., Hz (SW = 2 or 4 kHz) |
|---|---:|---:|
| $T_0^H$ | $\nu_{1H} - \nu_{1H}=0$ | 0 |
| $T_0^{HF}$ | $\pm(\nu_{1H} - \nu_{19F})=\pm 231$ | $\pm 231$ |
| $T_{\pm 1}^H$ | $\pm\nu_{1H}=\pm 3882$ | $\mp 118$ |
| $T_{\pm 1}^{HF}$ | $\pm\nu_{19F}=\pm 3651.5$ | $\mp 348.5$ |
| $T_{\pm 1}^{H-F}$ | $\pm(2\nu_{1H} - \nu_{19F})=\pm 4112.9$ | $\pm 112.9$ |
| $T_{\pm 2}^H$ | $\pm 2\nu_{1H}=\pm 7764.4$ | $\mp 235.6$ |
| $T_{\pm 2}^{HF}$ | $\pm(\nu_{1H} + \nu_{19F})=\pm 7533.7$ | $\mp 466.3$ |
| $T_{\pm 2}^{H-F}$ | $\pm(3\nu_{1H} - \nu_{19F})=\pm 7995.1$ | $\mp 4.9$ |
| $T^{H}_{\pm 3}$ | $\pm 3\nu_{1H}=\pm 11647$ | $\mp 353$ |
| $T^{HF}_{\pm 3}$ | $\pm(2\nu_{1H}+\nu_{19F})=\pm 11416$ | $\mp 584$ |
| $T^{H-F}_{\pm 3}$ | $\pm(4\nu_{1H}-\nu_{19F})=\pm 11877.4$ | $\mp 122.6$ |
| $T^{H}_{\pm 4}$ | $\pm 4\nu_{1H}=\pm 15529$ | $\mp 471$ |
| $T^{HF}_{\pm 4}$ | $\pm(3\nu_{1H}+\nu_{19F})=\pm 15298$ | $\mp 702$ |
| $T^{HF}_{\pm 5}$ | $\pm(4\nu_{1H}+\nu_{19F})=\pm 19180$ | $\mp 820$ |